\documentclass[10pt]{article}
\usepackage{amsmath}
\usepackage{graphicx}
\usepackage{tabularx}
\usepackage{epstopdf}
\usepackage{color}
\usepackage[a4paper,left=2cm,right=2cm,top=2cm,bottom=2cm]{geometry}

\title{A Novel Metamaterial-Inspired RF-coil for Preclinical\\ Dual-Nuclei MRI}

\author{
  A. Hurshkainen, A. Nikulin, I. Melchakova, P. Belov, S. Glybovski\\
  Department of Nanophotonics and Metamaterials, \\
  ITMO University, 197101 St. Petersburg, Russia
  \\
  \texttt{a.hurshkainen@metalab.ifmo.ru}
  \and
  E. Georget, B. Larrat\\
  DRF/I2BM/Neurospin/UNIRS, 91191 Gif-sur-Yvette Cedex, France\\
  \and
  D. Berrahou, L. Neves, P. Sabouroux, S. Enoch, R. Abdeddaim \\
  Aix Marseille University, CNRS, Centrale Marseille, Institut Fresnel,  F-13013 Marseille, France\\
}

\begin{document}

\maketitle

\thispagestyle{empty}

\begin{abstract}
In this paper we propose, design and test a new dual-nuclei RF-coil inspired by wire metamaterial structures. The coil operates due to resonant excitation of hybridized eigenmodes in multimode flat periodic structures comprising several coupled thin metal strips. It was shown that the field distribution of the coil (i.e. penetration depth) can be controlled independently at two different Larmor frequencies by selecting a proper eigenmode in each of two mutually orthogonal periodic structures. The proposed coil requires no lumped capacitors for tuning and matching. 
In order to demonstrate the performance of the new design, an experimental preclinical coil for $^{19}$F/$^{1}$H imaging of small animals at 7.05T was engineered and tested on a homogeneous liquid phantom and \textit{in-vivo}. The presented results demonstrate that the coil was well tuned and matched simultaneously at two Larmor frequencies and capable of image acquisition with both the nuclei reaching large homogeneity area along with a sufficient signal-to-noise ratio. In an \textit{in-vivo} experiment it has been shown that without retuning the setup it was possible to obtain anatomical $^{1}$H images of a mouse under anesthesia consecutively with $^{19}$F images of a tiny tube filled with a fluorine-containing liquid and attached to the body of the mouse.
\end{abstract}

\section*{Introduction}

Magnetic resonance provides unique instrumentation for modern biomedical studies. In clinical and preclinical applications radiofrequency (RF) coils play an important role of antennas exciting spins in a subject under applied strong static magnetic field with a desired flip angle and receiving back weak echo signals at the Larmor frequency. Thus a signal-to-noise (SNR) ratio of images, strongly depending on electromagnetic properties of  RF coils, becomes particularly important if a density of an investigated nucleus in the sample is low. 

In multi-nuclei studies it is possible to retrieve additional imaging and/or spectroscopy information using the magnetic resonance of several nuclei of interest typically taking place at different Larmor frequencies for the same magnet field $B_{0}$. The Larmor frequency of a nucleus is determined by its gyromagnetic ratio $\gamma$ as $f_{\text{L}}=\gamma{\cdot}B_0$.
The corresponding RF-coils suitable for multi-nuclei MRI must operate at all desirable Larmor frequencies simultaneously given that the input impedance is matched to the transceiver channel(s) and the RF-field distribution in a subject ensures maximum $B_1^{+}$ per unit power efficiency in transmission over the region of interest. The mentioned properties also ensure a high SNR of the same coil in reception improving imaging quality and resolution of measured molecular spectra.

Preclinical studies impose additional limitations to RF-coil designs. The dedicated coils must be compatible with other equipment required for biomedical experiment, such as excitation systems and sample beds. Another challenge comes from the electromagnetic environment of preclinical scanners, where RF-coils must operate inside an electrically narrow shielded tunnel. Due to a close proximity of a metal shield and electrically small dimensions, preclinical RF coils are inefficient, especially in ultra-high field small-animal imaging.
Finally, RF-coils are typically very sensitive to variation of subject's electric properties. 
In contrast to large coils (for instance, those used for clinical applications), where the noise from the subject dominates, in the so-called mid-range preclinical coils the subject noise becomes comparable with the intrinsic noise of the coil \cite{CoilReview}. The last fact results from power losses in the coil's components and strongly depends on the coil design and quality of its materials.  

RF-coils for small-animal imaging and spectroscopy at the fields 3--17 T are typically surface coils implemented as planar/conformal loops or volumetric coils (e.g. solenoids or bird-cage resonators) \cite{Naritomi1987611,Atalar1996596,Muftuler200239}. The most of conventional loop RF-coil designs have electrically small dimensions. When excited by an RF-cables, they exhibit inductive input impedance and, therefore, need to be tuned and matched at the desired Larmor frequencies using matching circuits with multiple on-chip capacitors \cite{CoilReview}. Smaller coil and bore dimensions require higher capacitance for tuning and matching which is accompanied by increasing losses and reducing SNR with increasing Larmor frequencies. Small-animal coils for scanning at two Larmor frequencies usually contain a tuning and matching circuit with at least two variable capacitors. A general dual-frequency matching strategy based on lumped capacitors has been proposed and validated in application to several different coils for the dual-nuclei $^{19}$F$/^{1}$H imaging at 4.7 T \cite{DualFH} (188 MHz and 200 MHz correspondingly). Lumped-element dual-frequency circuits advantageously provide similar RF-field patterns at both Larmor frequencies \cite{DualFH}. Also it has been proposed \cite{DualHNa} to use decoupled separate loops plugged to two separate channels of a 4 T MR system for scanning at $^1$H and $^{23}$Na (170 MHz and 45 MHz correspondingly). Again, both the channels were tuned and matched using variable capacitors.
In general, at least four independent parameters must be presented in the schematic to meet the tuning and matching conditions at two separate frequencies with no limitation of a quality factor. In practice, the capacitors of a circuit are controllable from outside of a bore using long dielectric rotating fixtures. The capacitors introduce losses due to strong localization of electric field increasing intrinsic noise, so that the requirements to their quality factors are very strong.

In this work we have proposed, studied and experimentally validated a novel dual-nuclei coil design for preclinical imaging. The proposed coil being self-resonant enables impedance matching at multiple frequencies due to resonant excitation of eigenmodes in multimode metamaterial-inspired structures of periodic metal strips. Noticeably, no lumped elements are required for operation of the coil. For the on-bench and MRI experiments we have manufactured the coil and tested it in $^{19}$F/$^{1}$H imaging on a phantom and \textit{in-vivo} with a small animal at 7.05 T, corresponding to the Larmor frequencies of 282.6 and 300.1 MHz with gyromagnetic constants of the nuclei: $\gamma(^{1}$H$)=42.58$ MHz/T, $\gamma(^{19}$F$)=40.05$ MHz/T. 

\section*{Results}
\begin{figure}[t]
  \centering
  \includegraphics[width=1\linewidth]{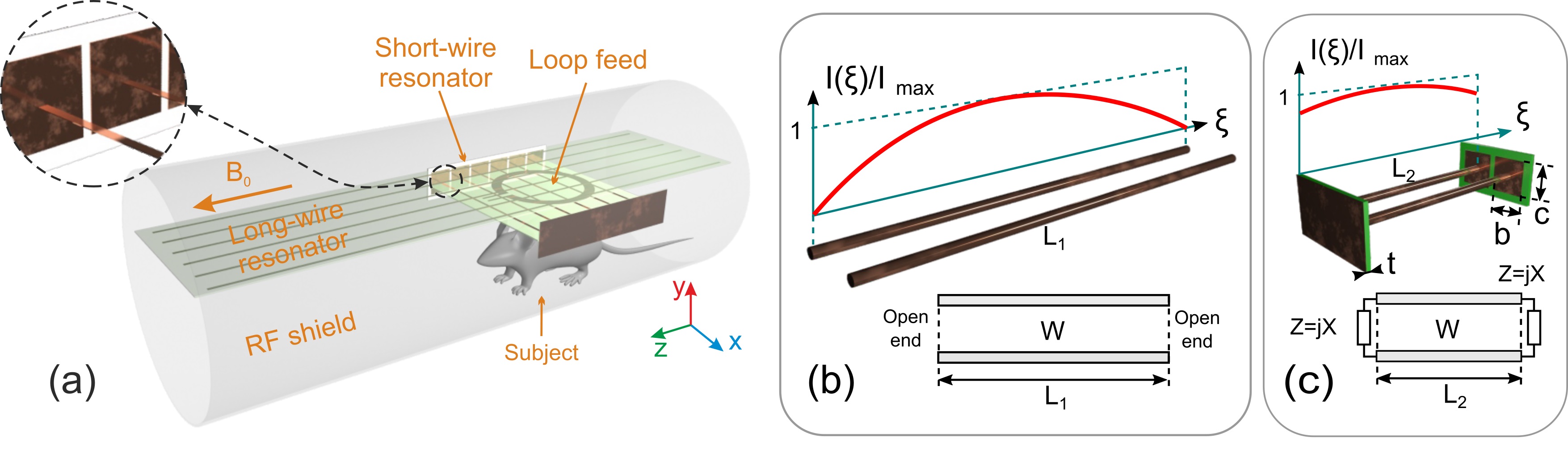}
  \caption{Proposed dual-nuclei RF-coil, comprising loop feed and two wire metamaterial-inspired resonators, with subject inside RF-shield of MRI (a); illustration of the miniaturization principle for the resonator of two metal wires: current distributions and equivalent circuits of the resonators with two half-wavelength (b) and shortened (c) parallel wires.}
  \label{Fig_ant}
\end{figure}

The proposed dual-frequency RF-coil operates due to excitation of eigenmodes in two mutually orthogonal periodic arrays of parallel thin metal strips by an external non-resonant circular loop feed connected to a single coax cable. These two wire metamaterial-inspired resonators are referred to as the long-wire resonator (with strips in parallel to the direction $z$ of the static field $B_0$ of the magnet) and the short-wire resonator (with transverse strips with respect to $B_0$). Excitation of eigenmodes is possible due to inductive coupling of the loop feed to both the resonators. The coil geometry with a subject inside a preclinical MRI bore is depicted in Fig. \ref{Fig_ant}(a). The long-wire resonator consists of identical nearly half-wavelength at 300.1 MHz strips, while the short-wire resonator for 282.6 MHz is based on shortened, but capacitively loaded strips fitting preclinical bore dimensions. The capacitive loads connecting the adjacent strips of the short-wire resonator are rectangular copper patches of the dimensions $b \times c$ printed on a common grounded dielectric substrate of the thickness $t$. Such capacitive loading presents at both ends of the shortened strips. The patches loading the strips of the short-wire resonator are shown in the inset of Fig. \ref{Fig_ant}(a) while their connection to two adjacent strips is illustrated in Fig. \ref{Fig_ant}(c).
\subsection*{Eigenmodes of wire metamaterial-inspired resonators}
In order to demonstrate the RF-field patterns created separately by the long-wire and the short-wire metamaterial-inspired resonators of the coil at their resonant frequencies, we have performed a numerical study of their eigenmodes. This resulted in two sets of resonant frequencies and two sets of RF magnetic field distributions.
These electromagnetic properties are useful to describe the physical principles behind the operation of the proposed coil under realistic MRI conditions. Instead of one electric-dipole-mode resonance of a single wire, in the array of $N$ half-wavelength wires (the long-wire resonator) due to the inter-wire coupling, one can excite $N=6$ eigenmodes with different resonant frequencies (so-called hybridization effect). Each eigenmode has its individual RF-field pattern created by a unique combination of currents flowing in positive or negative directions along the strips. Despite the fact that all currents related to a certain eigenmode have different magnitudes and phases (0 or $\pi$ in the lossless case), their distribution along each strip is similar and cosinusoidal (like at the resonance of a single wire).

\begin{figure}
  \centering
  \includegraphics[width=0.8\linewidth]{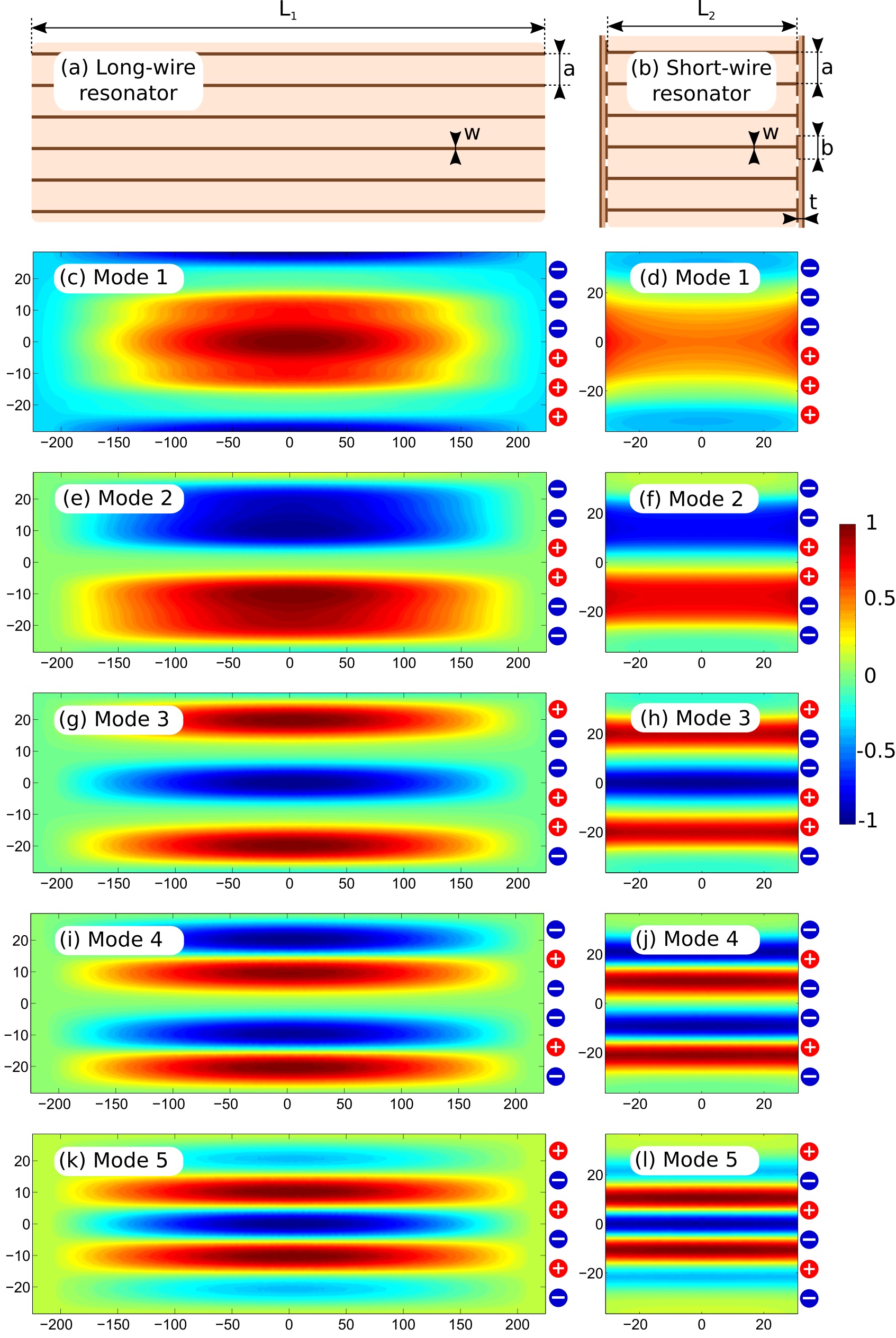}\\
\caption{Geometry and eigenmode H-field patterns (normal component with respect to the plane of strips), in a.u.: long-wire resonator (a,c,e,g,i,k) and short-wire resonator (b,d,f,h,j,l).}
\label{Fig_Eigenmode}
\end{figure}

In Fig. \ref{Fig_Eigenmode}(c,e,g,i,k) the calculated normalized distributions of $H_y$ field component (normal to the plane of strips) are presented for five of six available eigenmodes of the long-wire resonator suitable for preclinical MRI applications. Fig. \ref{Fig_Eigenmode}(a) describes the geometric parameters of the resonator. The field is shown in the plane parallel to the resonator, 5 mm away from the plane of strips.
Similar results, calculated for the resonator composed of $N=6$ shortened strips paired by connection through structural capacities (the short-wire resonator) are presented in the right column of Fig. \ref{Fig_Eigenmode}.
The short-wire resonator also supports six different eigenmodes, five of these are depicted in Fig. \ref{Fig_Eigenmode}(d,f,h,j,l). Fig. \ref{Fig_Eigenmode}(b) shows the geometric properties of the short-wire resonator. 

The numerically calculated resonant frequencies of the studied modes are listed in Table \ref{TabFreq}. 

\begin{figure}[t]
\center
\includegraphics[width=0.75\linewidth]{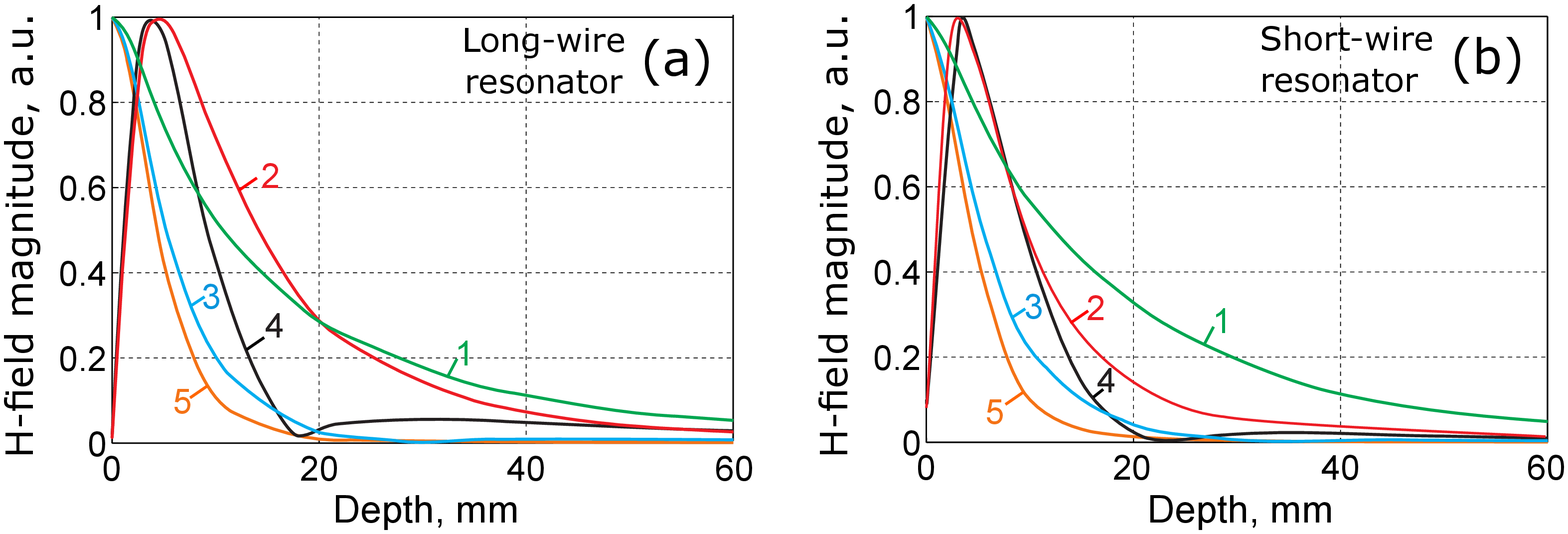}
\caption{Normalized magnetic field magnitude of the eigenmodes of the orders 1--5 depending on the distance away from the plane of strips: (a) long-wire resonator; (b) short-wire resonator.}
\label{Fig_depth}
\end{figure}
\begin{table}
\caption{Orders and simulated resonant frequencies of eigenmodes}
\centering
\begin{tabular}{|cc|cc|}
    \hline
    \multicolumn{2}{|c|}{Long-wire resonator} &  \multicolumn{2}{c|}{Short-wire resonator} \\
    Mode order & Frequency, MHz & Mode order & Frequency, MHz\\
    \hline
    1 & 322.4 & 1 & 264.4\\
    2 & 316.3 & 2 & 341.2\\  
    3 & 309.9 & 3 & 379.3\\  
    4 & 305.3 & 4 & 396.3\\ 
    5 & 302.5 & 5 & 408.6\\  
    \hline                  
\end{tabular}
\label{TabFreq}
\end{table}

In Fig. \ref{Fig_depth} the dependence of the normalized H-field magnitude as a function of the distance away from the plane of strips (the depth) is illustrated for the long-wire resonator (a) and for the short-wire resonator (b). The observation point in each case is located in front of the center of the resonator. 

\subsection*{Design and simulation of RF-coil}
\begin{figure}
\center
\includegraphics[width=0.8\linewidth]{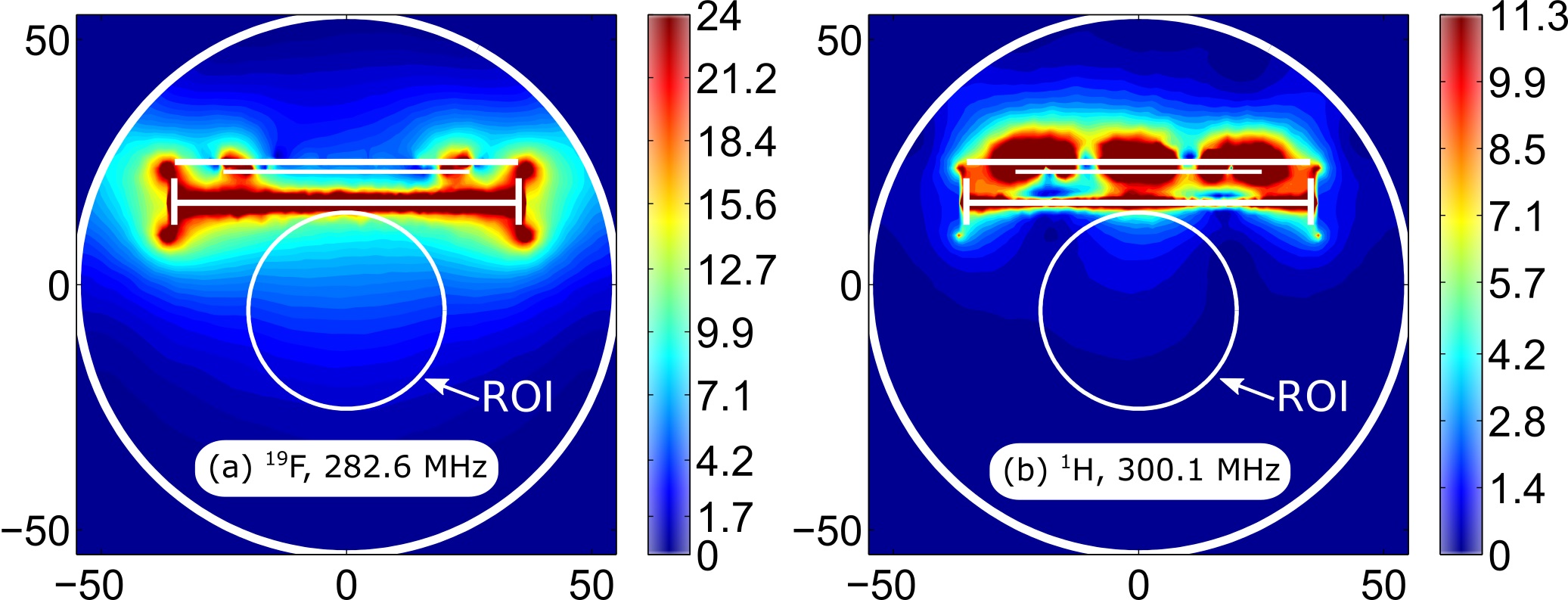}
\caption{Simulated distributions of $|B_1^{+}|$ for 0.5 W accepted power, $\mu\text{T}/\sqrt{\text{W}}$, in central transverse plane: 282.6 MHz (a) and 300.1 MHz (b).}
\label{Fig_transverse}
\end{figure}

The proposed dual-nuclei RF-coil as a whole combines together the long-wire and the short-wire resonators studied in the previous subsection into the same design. Strips of the two resonators are positioned in two parallel planes as shown in Fig. \ref{Fig_ant}(a). In the third parallel plane, a feeding non-resonant circular loop is located which is inductively coupled to both the resonators and is connected to a single coaxial cable connecting the coil to the transceiver at both Larmor frequencies. The main tuning parameters are the lengths $L_1$ and $L_2$ of strips, while the disposition of the loop feed with respect to the resonators is mostly responsible for matching. Selection of a proper eigenmode at each Larmor frequency allows to control the penetration depth. We have performed a full-wave numerical simulation of the assembled coil in the presence of a phantom (equivalent of a scanned subject) and the tunnel of MRI. In order to demonstrate that desired eigenmodes can be selected and adopted for operation of our coil at the given nuclei, we have tuned the mode 3 of the long-wire resonator (see Fig. \ref{Fig_Eigenmode}(g)) to the Larmor frequency of $^1$H and the mode 1 of the short-wire resonator (see Fig. \ref{Fig_Eigenmode}(d))) to the Larmor frequency of $^{19}$F. Due to the symmetry of their H-field distributions, both of these modes can be coupled to the circular loop located over the center of both resonators. Fig. \ref{Fig_s11,field}(a) shows the simulated reflection coefficient (red dotted curve) from the feeding point of the RF-coil in the split of the loop with respect to 50 Ohm cable impedance. The calculated normal magnetic field component distributions created by the RF-coil are shown for $^{19}$F and $^{1}H$ Larmor frequencies in Fig. \ref{Fig_s11,field}(b) and (c) correspondingly. Both the observation planes were chosen 5 mm away from the planes of the resonating strips. 
\begin{figure}
  \centering
  \includegraphics[width=1\linewidth]{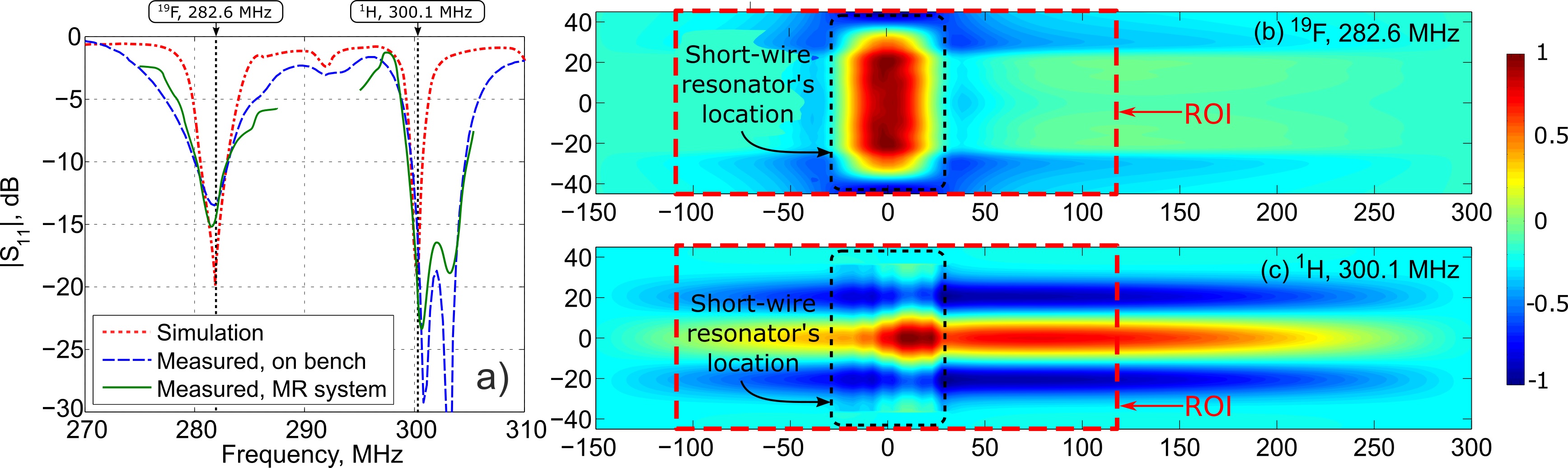}\\
  \caption{Simulated and measured values of the reflection coefficient $S_{11}$ of the proposed coil for $^1$H/$^{19}$F imaging vs. frequency (a) and simulated normal magnetic field component (a.u.) in vicinity of the assembled RF-coil: 282.6 MHz (b) and 300.1 MHz (c).}
  \label{Fig_s11,field}
\end{figure}
Fig. \ref{Fig_transverse} presents the distributions of the right-handed circularly polarized component $B_1^{+}$, which is responsible for excitation of spins. $B_1^{+}$ is plotted in the central transverse plane of the MRI bore and the field values correspond to the accepted power of 0.5W. 

\subsection*{MRI experiments on phantom and \textit{in-vivo}}

\begin{figure}[t]
\center
\includegraphics[width=1\linewidth]{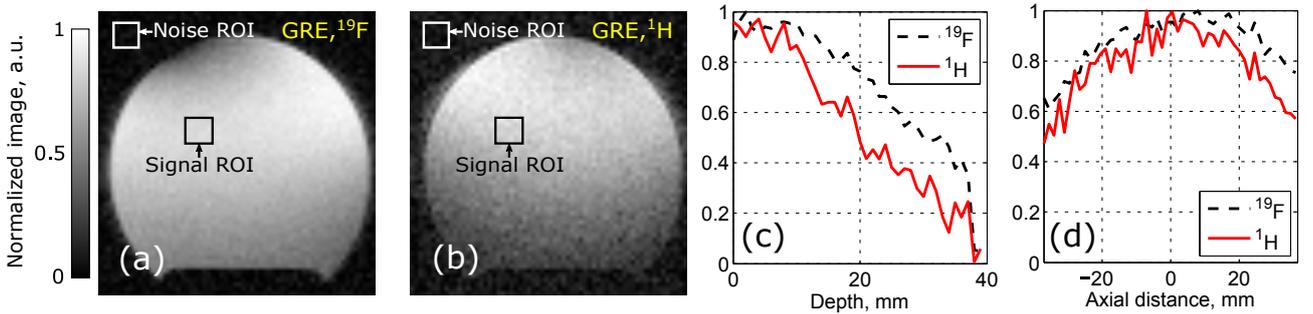}
\caption{Reconstructed images acquired with Bruker PharmaScan 7T of cylindrical liquid phantom (mixed 60\% 2-2-trifluorethanol and 40\% water) with $T_{1}$-weighed gradient echo sequence: $^{19}$F (a); $^{1}$H (b); normalized image profiles vs. depth to the phantom in the central transverse plane (c); normalized image profiles vs. axial distance along central axis of the phantom (20 mm depth) (d).}
\label{Fig_images}
\end{figure}

In order to experimentally test the proposed coil, two measurements under realistic MRI conditions have been performed. In the first one we used a homogeneous liquid phantom as a scanned subject to measure the S-parameters of the manufactured coil and acquire MRI images at the two nuclei of interest. In the second (\textit{in-vivo}) experiment we used a mouse under anesthesia for imaging at the same nuclei. The images of the cylindrical homogeneous phantom obtained for $^{19}$F and $^{1}$H without replacing the coil and the phantom are presented in Fig. \ref{Fig_images}. The measured SNR was calculated from the images as the ratio between the signal average and the noise standard deviation picked in the corresponding ROI shown in Fig. \ref{Fig_images}(a,b). For the hydrogen, the SNR in the ROI was 39 while for the fluorine SNR was 63. 

Additionally the decay of the image signal (in normalized image levels) at the two Larmor frequencies is illustrated in Fig. \ref{Fig_images}(c-d). The signal as the function of the depth in the phantom in the central axial plane is given in Fig. \ref{Fig_images}(c), while the signal depending on the axial distance along the central axis of the phantom (20 mm depth) is presented in Fig. \ref{Fig_images}(d).

\begin{figure*}
\center
\includegraphics[width=1\linewidth]{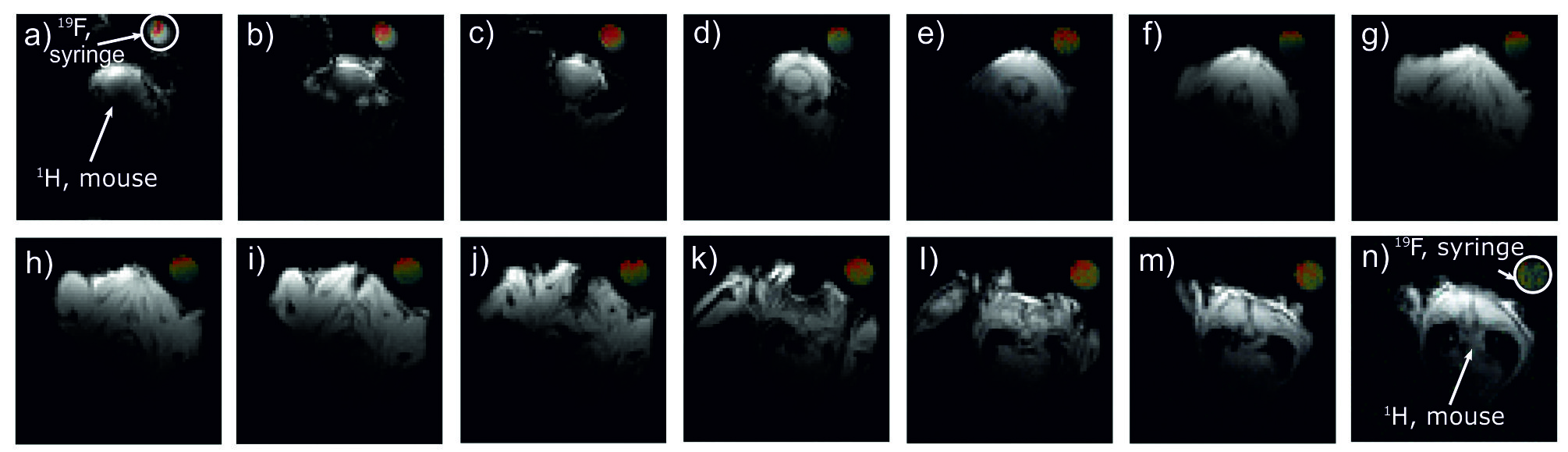}
\caption{\textit{In-vivo} anatomic $T_{1}$-weighted axial-plane images of mouse in 14 different slices acquired using the proposed dual-nuclei coil.}
\label{Axial_mouse}
\end{figure*}

\begin{figure}[t]
\center
\includegraphics[width=0.5\linewidth]{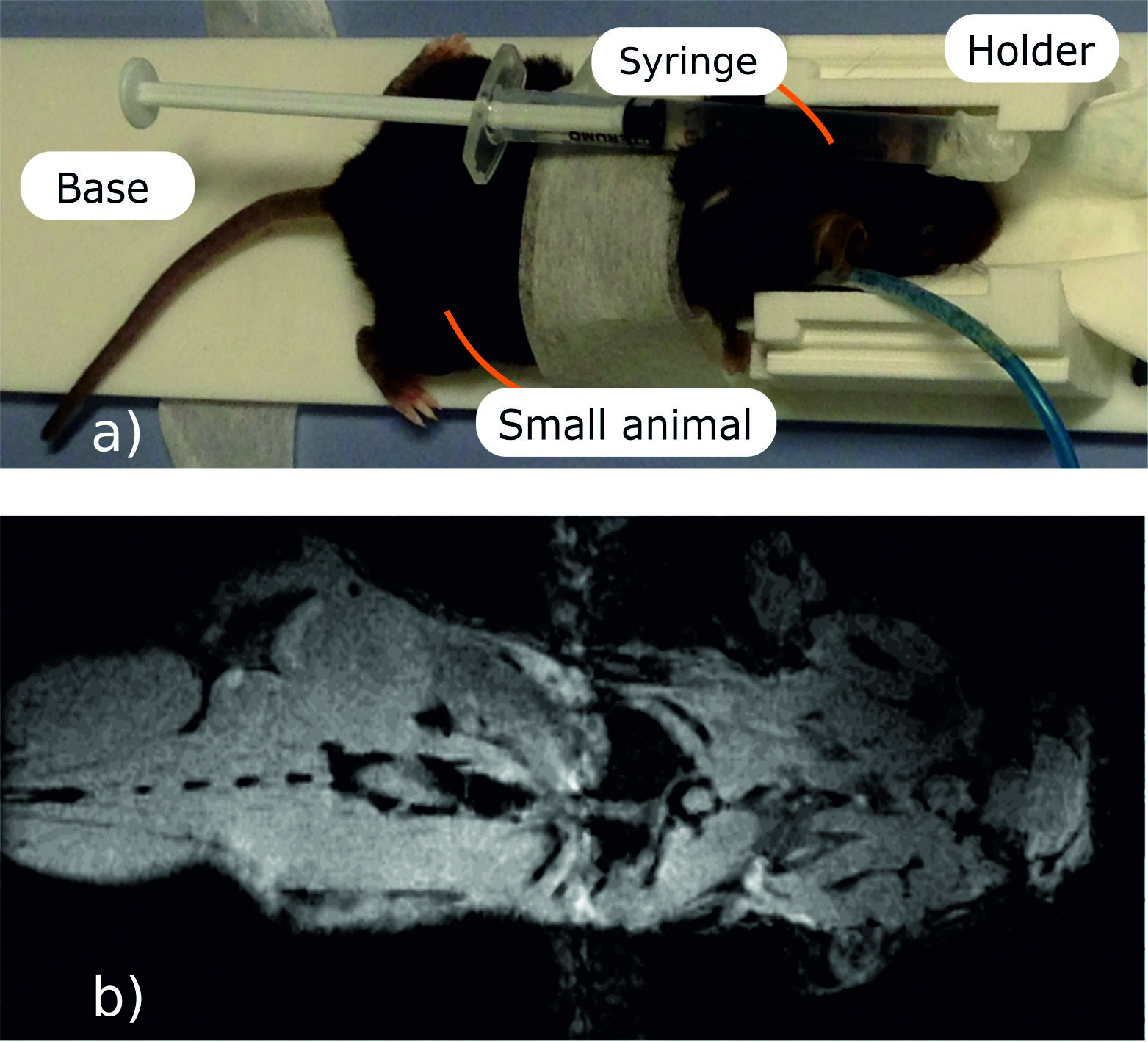}
\caption{\textit{In-vivo} setup including a mouse under anesthesia and a syringe containing fluorine compound (a) and anatomic $T_{1}$-weighted coronal-plane image of mouse acquired using the manufactured dual-nuclei coil (b).}
\label{mouse}
\end{figure}

The goal of the \textit{in-vivo} trials with the proposed coil was to demonstrate that the latter was capable of dual-nuclei imaging without replacing and retuning the setup. In the corresponding setup, a small syringe containing fluorine compound was attached to a mouse under anesthesia, as shown in Fig. \ref{mouse}(a). In Fig. \ref{Axial_mouse} images for the 14 adjacent transverse slices are displayed in a  \textit{gray} color scale showing the anatomy of the mouse body as well as the water inside the syringe. The corresponding images acquired at $^{19}$F are overlaid with the $^{1}$H images and mapped with a \textit{jet} color scale. As expected, only the syringe filled with fluorine compound gives signal.

An example of coronal $^{1}$H image of the whole mouse body is given in Fig. \ref{mouse}(b). No fluorine is present in this plane covering only mouse body. The coil shows capability to acquire anatomical images in mice with sufficient SNR and resolution. It has to be noted that the length of available imaging seems extended along the axis of the MRI as compared with a loop coil alone. 

The example of an anatomical coronal $^{1}$H image of the whole mouse body is given in Fig. \ref{mouse}(b).

\section*{Discussion}

An array of the given number $N$ of half-wavelength wires supports $N$ eigenmodes with different resonant frequencies. These resonances result from hybridization of the electric-dipole mode resonance of a single half-wavelength wire. The hybridization effect is due to interaction between identical coupled resonant elements of an array. The hybridization effect in periodic wire structure may lead to novel designs of RF-coils for MRI. Thus, the excitation of hybridized eigenmodes in arrays of coaxial resonators has been used in TEM-coils \cite{Avd}. Exciting such modes in resonant metasurfaces it is possible to locally improve SNR of an external RF-coil in 1.5T MRI. Recently \cite{Slobozhanyuk} it was shown that low-profile wire metamaterial-inspired resonators filled with water can serve as artificial pads locally increasing SNR of body coils. Also, it has been proposed to use the hybridized eigenmodes excited in a system of four $\lambda/2$-wires were used in a novel volumetric preclinical RF-coil for 7T \cite{Redha}.

Among the hybridized eigenmodes of the long-wire and the short-wire resonators, there is one with all the strip electric currents flowing in-phase. This extensively radiating mode has similar RF-field as the dipole mode of a single resonant wire. Since this mode is highly sensitive to the presence of an RF-shield, it has no practical interest in preclinical coils, but may be interesting for body imaging at 7T \cite{Raaijmakers_dipole}. All other eigenmodes necessarily have some out of phase currents pairs. A pair of currents with opposite phases produces a magnetic field localization nearby, which can provide an appropriate filling factor of a sample with dimensions comparable to the distance between the currents of the pair. In other words, for higher-order modes the H-field decays faster as function of the distance from the resonator in the normal direction. This is the common property of both the considered resonators of periodic strips: the long-wire and the short-wire resonators. 
 
For the long-wire resonator (left column in Fig. \ref{Fig_Eigenmode}), all the modes have a cosine-like field distribution of the normal magnetic field component in the longitudinal direction (along the strips). At the same time, the distribution has a standing-wave shape with 1–5 lobes in the transverse direction with respect to the strips. The number of lobes is related to the order of the mode. Note that in Fig. \ref{Fig_Eigenmode} the eigenmodes are sorted based on the number of standing-wave lobes (the order of the standing-wave resonance). The positive (0) or negative ($\pi$) phases of strip currents producing the mode patterns are indicated in each pattern by ”+” or ”-” symbols to the right from the field color plots. One can check that the most homogeneous field pattern (mode 1, Fig. \ref{Fig_Eigenmode}(c)) corresponds to the highest resonant frequency. In contrast, the mode 5 having 5 standing-wave lobes resonates at the lowest frequency. All the simulated resonant frequencies for the modes 1–5 are compared for the two resonators in Table \ref{TabFreq}.
 
The modes calculated for the short-wire resonator having resonant frequencies approximately in the same range around 300 MHz (see Table \ref{TabFreq}), are depicted in the right column in Fig. \ref{Fig_Eigenmode}. One can observe that the magnetic field distributions are similar to that of the long-wire resonator, especially in terms of the standing-wave profile in the direction orthogonal to the strips. However, the field in the direction of strips is almost constant due to the presence of structural capacitive loads at the ends of the strips. In fact, for the short-wire resonator this current profile along the strips is given by the same cosine-like function as for the long-wire resonator. The difference is that this profile is cropped by the electrically small strip lengths. The latter is illustrated by the current distributions in Fig. \ref{Fig_ant}(b,c). The other difference of the short-wire resonator is the inverse order of resonant frequencies with respect to the long-wire resonator. This inverse order of modes can be explained by much higher contribution of the capacitive coupling between adjacent strips for the short-wire resonator than for the long-wire one.  
 
In the considered two resonators the mode with the most homogeneous pattern of the normal H-field component is the mode 1 (Fig. \ref{Fig_Eigenmode}(c,d)). From Fig. \ref{Fig_depth}(a,b) one can see that this mode has the slowest decay of the field as the observation point goes away from the plane of strips. It has the lowest resonance frequency of 264.4 MHz for the short-wire resonator, while having the highest resonance frequency 322.4 MHz for the long-wire resonator. It can be observed that the field decays faster for higher number of lobes. For RF-coil application one can expect that the modes with better in-plane homogeneity of the normal H-field component show better penetration to a subject and higher $B_1^{+}$ per power efficiency at a depth. It should be noted that odd and even modes must be compared separately, as even modes have zero H-field at the center of the resonator. The highest penetration depth can be expected from the mode 1 (Fig. \ref{Fig_Eigenmode}(c,d)). The distribution of electric current phases of this mode provides that the field in the vicinity of the resonator is similar to the field of a long surface loop coil. If one aims to localize the field pattern in a subject’s surface layer, the mode with the most inhomogeneous normal magnetic field and current phases should be excited in the resonator (namely the mode 5, Fig. \ref{Fig_Eigenmode}(k,l)). This mode is equivalent to a set of 5 narrow and long loop surface coils, where each loop is one period in width and the neighboring loops are excited out-of-phase, resulting in low field penetration. 
 
In the both long-wire and short-wire resonators it is possible to excite a variety of eigenmodes, which differ by their resonant frequencies and field penetration depths. If the two resonators are combined in the same RF-coil, one can manage its field penetration to a subject at two different Larmor frequencies by selecting appropriate resonant eigenmodes (one selected mode from each of two resonators).

For a dual-nuclei MRI one needs to operate at two different Larmor frequencies, possibly with the same or different field patterns depending on the application. In the above discussed resonators, it could be possible to use two of N eigenmodes exciting them at two different Larmor frequencies. However, for neither the long-wire nor the short-wire resonator the resonant frequencies of two different eigenmodes cannot be tuned independently. Therefore, in the proposed coil we have combined into the same design the two resonators using just one eigenmode from each of them. Therefore, in the proposed design the short-wire resonator is responsible for $^{19}F$, while the long-wire resonator is responsible for $^{1}H$. The small overlap area of the two resonators in comparison to the area of the long-wire resonator minimizes their mutual coupling. On another hand, since the strips of the short-wire resonator are arranged perpendicularly to the $B_0$ direction and the typical bore diameter of a 7T preclinical bore is only 90 mm, the resonator must be considerably shortened in comparison to the half-wavelength. This can be achieved by capacitive connection in each pair of strips \cite{Mauritius} as shown in Fig. \ref{Fig_ant}(c). This self-resonant geometry comprising periodic metal wires and capacitive patches at their ends was inspired by mushroom high-impedance surfaces \cite{Sie99} and realized by connecting each printed shortened strip of the resonator end to a rectangular copper patch of the side lengths b = 9 and c = 9.5 mm. All the patches from both ends of the strips were printed on two separate Rogers 4003C grounded substrates of the thickness t = 0.508 mm (see Fig. \ref{Fig_ant}(c) and Fig. \ref{Fig_Eigenmode}(b)).
 
The idea of miniaturization by connecting adjacent strips through capacitive loads can be understood on the following example of the resonator with only two strips. The long-wire resonator with two strips of the width $w$ and the separation $a$ acts as a TEM-transmission-line segment with two open ends. Even if the strips are printed on a thin dielectric substrate, the initial resonant length of this resonator is approximately equal to $L_1=\lambda/2$, where $\lambda$ is the wavelength in free space (see Fig. \ref{Fig_ant}(b)). In contrast, in the short-wire resonator the parallel strips are connected at both ends to each other through two structural capacities, each one realized as a rectangular patch over a grounded substrate. The resonator together with its equivalent-circuit are shown in Fig. \ref{Fig_ant}(c). From the circuit, it is possible to derive the formula for the resonant length $L_2$ shortened due to the capacitive loads:
\begin{eqnarray}
L_2=\frac{\lambda}{2\pi} \left(\arctan \frac{2WX}{X^2-W^2}+\pi \right),~-\infty<X<-W; \label{L21}\\
L_2=\frac{\lambda}{2\pi} \arctan \frac{2WX}{X^2-W^2},~-W<X<0; \label{L22}
\end{eqnarray}
The structural load reactance $X$ can be calculated from a series connection of two similar capacities $C_{\text{patch}}$ between a patch and a ground plane:
\begin{equation}\label{ex}
X=-\frac{2}{\omega C_{\text{patch}}},
\end{equation}
where $\omega=2\pi f$ is the angular frequency and $C_{\text{patch}}$ is given by
\begin{equation}\label{ec}
C_{\text{patch}}=\frac{\varepsilon_r\varepsilon_0 b\cdot c}{t}.
\end{equation}
The formulas (1-2) can be used for estimation of the miniaturization factor due to the capacitive loading. In other words, one can calculate the ratio $L_1/L_2$ for the given geometric parameters of patches and strips. For the above considered geometric parameters $a$, $b$, $c$ and $w$ at 300 MHz using the analytical formula for the wave impedance W of an edge-coupled strip line \cite{Cohn55}, one can obtain $L_1/L_2=4.3$. However, in the real resonator configuration with six strips, the above formulas give only an approximation of the lengths of shortened strips. The certain values $L_1$ and $L_2$ for tuning at the given Larmor frequencies depend on the orders of selected eigenmodes and can be precisely predicted numerically. The certain numerically determined length of strips of the short-wire resonator was $L_2 = 72$ mm, while the length of the long-wire resonator was $L_1=434$ mm.
 
The whole proposed coil (see Fig. \ref{Fig_ant}(a)) has been numerically calculated in the realistic MRI setup (with a phantom and the RF-shield) to demonstrate its dual-frequency operation. For the two Larmor frequencies, corresponding to the nuclei of interest $^{19}F$ and $^{1}H$ we decided to select the following modes: the mode 1 of the short-wire resonator and the mode 3 of the long-wire resonator correspondingly. This choice was made to demonstrate the flexibility of the proposed design in terms of the field penetration depth for a given coverage of the sample's surface. Thus at 282.6 MHz we expected to reach high penetration depth due to the first mode, while having a strong surface localization at 300.1 MHz due to the third mode. Provided that the mentioned modes belong to different resonators it is easy to tune their resonances to the Larmor frequencies almost independently by varying the lengths of the corresponding strips. Matching at both frequencies was achieved by adjusting their positions with respect to a common small loop feed connected to a 50-Ohm port. The resulting simulated frequency curve of the reflection coefficient $S_{11}$ (dotted red line in Fig. \ref{Fig_s11,field}(a)) shows that indeed the dual-frequency tuning and matching is possible with the certain geometric parameters of the proposed coil ($|S_{11}|$ is much lower than -10 dB at both the Larmor frequencies) with no lumped capacitors required. The magnetic field distributions created by the coil at these frequencies are presented in Fig. \ref{Fig_s11,field}(b,c) and Fig. \ref{Fig_transverse}. As can be seen from the comparison of Fig. \ref{Fig_s11,field}(b) and Fig. \ref{Fig_s11,field}(c) with Fig. \ref{Fig_Eigenmode}(d) and \ref{Fig_Eigenmode}(g), the RF-field distributions obtained with the assembled coil are mostly determined by the desirable eigenmodes of the short-wire and the long-wire resonators. However, a small field distortion at both frequencies can be observed in the location where the two resonators overlap (i.e. the location of the smaller short-wire resonator). The distortion occurs due to an inductive coupling between the selected eigenmodes of the resonators. Comparing the distributions inside the phantom at the two Larmor frequencies (Fig. \ref{Fig_transverse}), it can be concluded that the field penetration at 282.6 due to the mode 1 of the short-wire resonator is indeed deeper than one of the long-wire resonator’s mode 3 at 300.1 MHz. This result is in qualitative agreement with the curves of Fig. \ref{Fig_depth}. It should be noted that from the field maps in Fig. \ref{Fig_transverse} one can clearly determine which resonator is responsible for excitation of the phantom at the Larmor frequencies. Therefore, as shown in the simulation, the proposed coil design allows tuning and impedance matching at two predefined Larmor frequencies using no lumped capacitive elements due to resonant excitation of the selected eigenmodes.
 
Despite of the orthogonal orientation of two resonators and a relatively small size of the short-wire resonator, the effect of their mutual coupling is still observable. Due to this effect and the influence of the RF-shield, the resonance frequencies of the selected modes moved from 309.9 MHz (mode 3 of the long-wire resonator) and 264.4 MHz (mode 1 of the short-wire resonator) to 300.1 and 282.6 MHz respectively. It should be noted that since the mutual coupling of the resonators is relatively weak, their resonant frequencies are mostly dependent on the length of their own strips. 
 
The same design as the above simulated has been manufactured and tested on the bench and in MRI scans. The two $S_{11}$ curves measured with a VNA on the bench (dashed blue curve in Fig. \ref{Fig_s11,field}(a)) and using the MR system (solid green curve in Fig. \ref{Fig_s11,field}(a)) demonstrate good agreement with the simulated data. The difference in $S_{11}$ levels at the two resonances and the presence of additional parasitic resonances can be explained by the influence of a realistic MR-system’s bore geometry and the gradient system configuration, which was not considered in simulations. However, the experimental results confirm sufficient matching ($S_{11}<-10$ dB at both the Larmor frequencies) required for dual-nuclei scanning.
 
The same coil setup with the phantom was used in the MRI scans. Comparing the obtained phantom images, one can conclude that the SNR decays with the depth in the phantom for $^{1}H$ much faster than for $^{19}F$, which agrees with the numerically calculated field plots in Fig. \ref{Fig_transverse}. The SNR of the hydrogen image is also lower since the distance to the long-wire resonator from the phantom is slightly larger than the distance to the short-wire resonator. In fact, one could replace the long-wire and the short-wire resonators by each other to improve SNR in the top surface layer of the phantom for $^1H$. On the top left of the $^{19}$F image the dark area can be observed. The origin of this effect is a high intensity of the $B_1^{+}$ in the close proximity of the short-wire resonator, which is disposed just above the phantom. Flip angle in this part of the phantom becomes large than the optimal one (close to 90 degrees) and it refers to a so-called overflip. The obtained phantom images clearly demonstrate that the field distributions created by the coil at two Larmor frequencies are almost similar in the axial direction of the MR-system, while the penetration depths are significantly different. This behavior of the coil results from the multi-mode nature of the periodic wire resonators and is determined by the selected eigenmode properties, as expected from simulations.
 
The obtained \textit{in-vivo} images presented in Fig. \ref{Axial_mouse} and \ref{mouse} demonstrate that the designed coil is capable of $^1H$ anatomic imaging of a top parts of a body of a small animal with a wide coverage area in the coronal plane. As was shown above the limited penetration at the Larmor frequency of $^1H$ is governed by the properties of the selected Mode 3 of the long-wire resonator. At the same time, the field of view of the coil is long in the $B_0$ direction covering all the body length of a small animal, which also goes along with theoretical predictions. The coil was also capable of $^{19}$F imaging of the syringe under test using the same setup without retuning the coil. The performed dual-nuclei imaging has an appropriate quality for various biomedical applications of $^{19}F/^{1}H$ MRI.

\section*{Methods}
\subsection*{Eigenmodes of wire metamaterial-inspired resonators}
In order to numerically calculate the resonant frequencies of the two wire metamaterial-inspired resonators of the proposed coil (see Fig. \ref{Fig_ant}(a)) and their eigenmode field distributions we have made simulations using the Eigenmode Solver in CST Microwave Studio 2016 commercial software. For the short-wire and the long-wire resonators the geometric parameters are illustrated in Fig. \ref{Fig_Eigenmode}(a,b). Both resonators are single-layer flat periodic arrays of $N=6$ thin printed copper strips of the width $w=1$ mm and periodicity $a=10$ mm. The substrate of the printed circuit board was 0.508-mm-thick Rogers 4003 laminate with permittivity $\varepsilon_r=3.38$ and the dielectric loss tangent 0.0027. The long-wire resonator depicted in Fig. \ref{Fig_Eigenmode}(a) has the length of strips $L_1=434$ mm, which is comparable to one half of the wavelength at 300 MHz. 
The long-wire resonator was modeled alone in the lossless approximation, without a subject and a feeding loop. For the short-wire resonator the same numerical eigenmode analysis was performed separately. The following geometric parameters of the short-wire resonator were set: $L_2=72$ mm, $b=9$ mm and $c=9.5$ mm. All the patches from both ends of the strips of the short-wire resonator were printed on two separate Rogers 4003 grounded substrates of the thickness t = 0.508 mm (see Fig. \ref{Fig_ant}(a) and Fig. \ref{Fig_Eigenmode}(b)).
In the simulation, the number of analyzed modes was set to 5 for both the resonators. The calculation domain was bounded by distant PEC walls from all sides. No phantom was included to the eigenmode simulation.

\subsection*{Design and simulation of RF-coil}
The whole proposed coil assembled of two metamaterial-inspired resonators and a small loop feed was simulated using Frequency Domain Solver in CST Microwave Studio 2016 together with a homogeneous phantom and a perfectly conducting cylindrical RF-shield of the diameter 90 mm. The long-wire resonator was located on top of the short one as depicted in Fig. \ref{Fig_ant}a, so that the separation between their PCBs was 9 mm. In between of the resonators a small inductively coupled feed was placed implemented as a flat annular copper ring of the width 4 mm and the external radius 20 mm, printed on 0.508-mm-thick Rogers 4003C substrate. The loop was fixed 7 mm away from the plane of the long strips and 2 mm away from the plane of the short strips. As a result, a three-layer stack of PCBs was composed operating as a dual-frequency surface coil. The coil produced an RF-field inside the tightly located cylindrical phantom with the diameter 40 mm, height 75 mm and material properties $\varepsilon_{\text{r,phantom}}=39$ and $\tan\delta=0.06$ S/m. In the simulation, the loop was driven by a lumped 50-Ohm port connected to its split, and the small loop itself excited one selected eigenmode in each of the two wire resonators. The parameters of the resonators were the same as preciously used for eigenmode analysis. The values $L_1$ and $L_2$ were chosen in a parametric sweep performing multiple field simulation. The optimization goal was to achieve the calculated $|S_{11}|$ well below -10 dB at both the considered Larmor frequencies. Impedance matching was reached by variation of the distances between the three above mentioned parallel PCBs of the RF-coil. Once tuning and matching is reached at the two desired Larmor frequencies, the circularly polarized RF field $B_1^{+}$ rotating in the axial plane with respect to the static field of the magnet corresponding to the accepted power of 0.5W was calculated using the template-based post-processing routine in CST Microwave Studio. 

\subsection*{On-bench measurements}

\begin{figure}
\center
\includegraphics[width=1\linewidth]{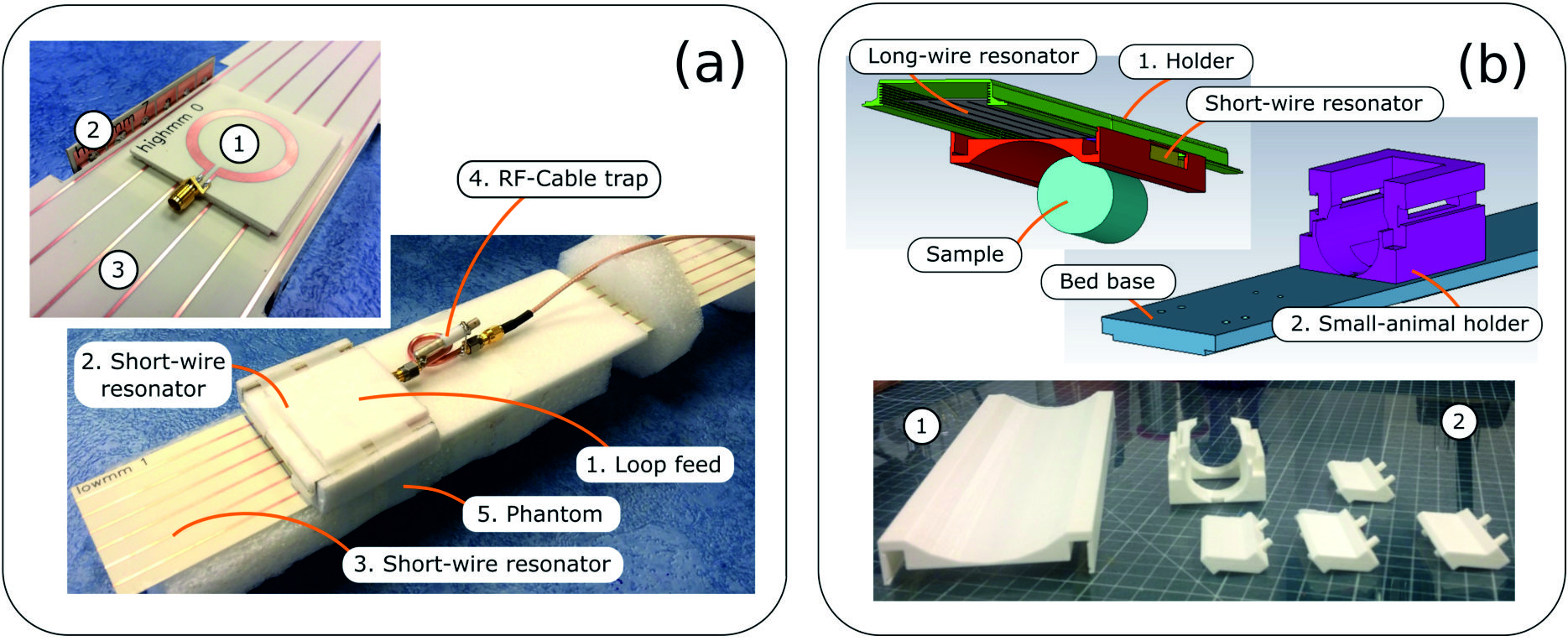}
\caption{Experimental RF-coil: (a) PCB parts including loop feed, short-wire resonator and long-wire resonator and their assembly with a spacer, phantom and RF-cable; (b) bed parts: holder for resonators and for small animal.}
\label{Fig_photo}
\end{figure}

For on-bench and MRI experiments the proposed coil was manufactured by manual assembling of separate PCB parts, shown in inset of Fig. \ref{Fig_photo}(a), based on a common specially designed 3D-printed bed. The bed included a holder supporting the antenna parts shown in Fig. \ref{Fig_photo}(a). The PCB parts in Fig. \ref{Fig_photo}(a) are labeled as: 1 --- loop feed with SMA connector, 2 --- short-wire resonator with soldered PCBs supporting capacitive patches, 3 --- long-wire resonator. The bed also includes the base supporting the coil and a removable holder for a scanned sample, i.e. a small animal (see Fig. \ref{Fig_photo}(b)). The whole bed fits the dimensions of a 90-mm diameter bore of a preclinical MRI. In Fig. \ref{Fig_photo}(a) one can also observe the feeding coax with the RF-cable trap. The parts of the bed were manufactured by 3D printing.

All the geometry parameters of the two resonators in the manufactured coil correspond to ones chosen in the previously discussed simulation.
The short-wire resonator was constructed of three 0.508-mm-thick Rogers 4003 PCBs: one supporting shortened copper strips and two grounded side PCBs with rectangular patches. All six strips from both ends were manually soldered to the corresponding patches printed on two side PCBs.
The long-wire resonator was represented only by a single Rogers 4003 PCB with strips without connection to any capacitive loads. The coil was fed with a coaxial 2-mm 50-Ohm cable connected to the printed non-resonant circular loop through the cable trap with a cable ring shunted by a variable capacitor.
The homogeneous phantom was represented by a cylindrical plastic can with the diameter 40 mm and height 75 mm filled with the solute of 60\% 2-2-trifluorethanol and 40\% water. The permittivity of 39 and the dielectric loss tangent 0.06 were measured for this liquid in the frequency range 100-400 MHz using the calibrated coaxial line section \textit{EpsiMu} connected to the vector network analyzer (VNA) Anritsu MS2036C, and the obtained values were used in the performed numerical simulations.

In order to ensure proper tuning and impedance matching the reflection coefficient $|S_{11}|$ of the coil was measured first with the vector network analyzer (VNA) Anritsu MS2036C connected through a long calibrated cable to the coil located inside the bore of Bruker PharmaScan 7T MR system. The result is shown in Fig. \ref{Fig_s11,field}a with a dashed blue curve. 

\subsection*{MRI experiments with phantom and \textit{in-vivo}}

In order to validate the obtained tuning and impedance matching of the experimental coil, $|S_{11}|$ was again measured by means of the MR-system within its two operational bands (represented in Fig. \ref{Fig_s11,field}a with the solid green curves).

The coil was tested by imaging of the phantom by scanning in the MR-system using the pulse sequence $T_{1}$-weighed gradient echo (FLASH) TR/TE=2000/2.4 ms with an isotropic voxel $0.7\times 0.7 \times 0.7$ mm$^3$.

\textit{In-vivo} mouse acquisition was performed in accordance with European Union and French laws on animal experiment (project authorization number: 12-058, site authorization number: B-91-272-01). One mouse was placed under anesthesia (isoflurane 1-2\%) in front of the center of the feeding loop.  A 1 mL syringe filled with a mixture of 60\% 2-2-trifluoroethanol and 40\% water was attached to the back of the mouse as shown in Fig. \ref{mouse}(a). The mouse was positioned under the RF-coil’s resonators in front of the loop coil.

Two sets of images were acquired sequentially at Larmor frequencies of 282.6 MHz for $^{1}$H and 300.1 MHz for $^{19}$F without need for any intervention on the coil in between acquisitions. 14 transverse slices were acquired using a $T_{1}$-weighted gradient echo sequence (FLASH) TR/TE=2000/2 ms with a voxel size $0.5\times 0.5$ mm$^2$ and the slice thickness of 2 mm (see the obtained whole-body image in Fig. \ref{mouse}(b)).

In addition to transverse images, coronal images at higher resolution were also acquired by means of a $T_{1}$-weighed gradient echo sequence (FLASH) TR/TE=2000/4.6 ms with the spatial resolution of $0.25\times 0.25$ mm$^2$ and the slice thickness of 0.5 mm (see the obtained whole-body image in Fig. \ref{mouse}(b)).

\section*{Conclusion}
In this work, a new design of a dual-nuclei preclinical RF-coil was proposed and characterized. The operational principle of the coil is based on resonant excitation of eigenmodes in a pair of multi-mode wire metamaterial-inspired resonators. It was numerically and experimentally shown that by proper selection of the excited eigenmodes one can control the penetration depth in a subject, i.e. to excite efficiently the whole body or just a particular surface of a small animal. The resonant frequencies of the selected modes can be tuned by geometrical properties of the resonators of the coil, while impedance matching is determined by mutual positions of the resonators and the feeding annular loop. It was experimentally shown by measurements on a phantom and textit{in-vivo} that the proposed coil is suitable for dual-nuclei $^{19}$F/$^{1}$H imaging at 7T with good homogeneity and SNR sufficient for small-animal biomedical studies. The proposed coil design is also compatible with multi-nuclei MRI  applications using other nuclei (e.g. $^{23}$Na, $^{31}$P etc.) since both the short and the long-wire resonators can be tuned to various Larmor frequencies by adjusting their length and the parameters of patches (providing the distributed load capacity for the short-wire resonator). Importantly, the proposed self-resonant design is cheap as it contains no variable non-magnetic capacitors for tuning and matching and can be constructed of only several PCB parts movable against each other.


\section*{Acknowledgment}

This project has received funding from the European Union's Horizon 2020 research and innovation program under grant agreement No 736937. This work was supported by the Ministry of Education and Science of the Russian Federation (Zadanie No. 3.2465.2017/4.6).
The authors would like to thank Prof. Constantin Simovski for useful discussions.

\section*{Additional information}
\textbf{Competing interests}: The authors declare that they have no competing interests.

\end{document}